# The Long-Range Memory and the Fractal Dimension: a Case Study for Alcântara

Cleber Souza Corrêa[1], Daniel Andrade Schuch[2], Antonio Paulo de Queiroz[1], Gilberto Fisch[1], Felipe do Nascimento Corrêa[1], Mariane Mendes Coutinho[1]

**ABSTRACT:** This study aimed to analyze the time series behavior of the Southern Oscillation Index through techniques using Fast Fourier Transform, computing the autocorrelation function, and the calculation of the Hurst coefficient. The methodology of Hurst Exponent calculation uses different lags, which are computed in the time series of Southern Oscillation Index. The persistent behavior in the time series can be characterized by calculating the Hurst Exponent, seeking for more behavioral information, as the existence of persistence and/or terms of long-range memory in the series. The results show a persistence of the climate in terms of long-memory Southern Oscillation Index time series, which can help to understand a complex dynamic behavior in climate effects at global-scale level and specifically its influence in northeastern Brazil, in the region of the Alcântara Launch Center. The R package tseriesChaos was used in the analysis of the Southern Oscillation Index time series, estimating the largest Lyapunov exponent, which indicates the existence of chaotic behavior in time series. The resampling technique was used in a permutation test between the surface wind data in the São Luís airport, Maranhão State, and the Southern Oscillation Index. The permutation test results showed that the time series of monthly average wind speed in the São Luís airport is correlated with the variability of Southern Oscillation Index, statistically correlated to the confidence interval at the 5% level. The results showed the possibility of using autoregressive models to represent average variable meteorology in the behavior analysis as well as trends in the climate, more specifically a possible climatic influence of El Niño-Southern Oscillation in wind strength in the Alcântara Launch Center.

# INTRODUCTION

Sea surface temperature conditions in the tropical Pacific are important drivers of the atmospheric circulation and can have a major influence on the global climate. The El Niño-Southern Oscillation (ENSO) is a key component of the climate system (Capotondi 2013) and it is important to understand possible changes in its variability, which may be due to natural processes, such as the decadal variability of climate or anthropogenic effects.

The dynamic behavior of ENSO is very complex, and it is difficult to predict its time series variations. The Southern Oscillation Index (SOI) time series signal is composed of a set of geophysical forcing with different temporal scales, some of long time as the cycles of solar activity and/or the interaction atmosphere-ocean, involving a planetary scale. The Southern Oscillation is a seesaw in surface air pressure between the tropical eastern and the western waters of the Pacific Ocean. SOI is calculated as the standardized anomaly of the surface air pressure difference between Tahiti, in the Pacific Ocean, and Darwin, Australia, in the Indian Ocean.

The Southern Oscillation is a dynamic coupled ocean/atmosphere process that influences the planetary scale. It characterizes certain phases: when the situation is positive, it means that there are higher pressure values in Tahiti and lower ones in Darwin (La Niña); the negative phase is when the pressure values are lower in Tahiti and higher in Darwin (El Niño); and in the neutral phase there are no significant values. ENSO is a planetary-scale phenomenon that occurs naturally in the tropical Pacific with global and highly-relevant impacts, affecting greatly the human society. El Niño refers to the heating above normal in the tropical Pacific Ocean, which occurs with a frequency of 2–7 years. Its opposite phase, when the tropical Pacific Ocean is colder than normal, is known as La Niña. These sea surface temperature changes affect the weather with values above or below the climatological ones.

Capotondi *et al.* (2015) improved the determination and understanding of ENSO, its predictability and the existence of teleconnections. However, given the complexity and impacts of ENSO, a better analysis of the spatial patterns and their evolution is required. Newman (2007) concludes that the long-term predictability exists; however, due to complex characteristics of the dynamic system in the planetary circulation, the canonical ENSO and the Pacific Decadal Oscillation (PDO) patterns, the influence of the interaction of these systems creates a complex structure, which consists of 2 stationary eigenmodes that are weaklier damped. These eigenmodes can represent different effects: the canonical ENSO (Barnston and Ropelewski 1992; Gershunov and Barnett 1998; Takahashi *et al.* 2011) between the years 1900–2002 may represent the influence of anthropogenic nature and multidecadal fluctuations of a pattern that is potentially a natural decadal variability. The predictability of these stationary eigenmodes is significantly increased by the coupling of these eigenmodes between the North Pacific region and its tropical oceanic part.

Gershunov and Barnett (1998) showed that the PDO has a modulatory effect on climatic patterns creating ENSO events. As a result, there are important characteristics to be observed: when the phase is El Niño, that stage is likely to be more intense when the PDO is highly positive; on the other hand, the La Niña phase is more intense when the PDO is strongly negative. This behavior does not mean that the PDO physically controls the dynamic ENSO, but the resulting climatic patterns show the interaction between them.

The influence of the Southern Oscillation in the northeastern Brazil climate has already been well-studied in the literature (*e.g.* Enfield and Mayer 1997; Uvo *et al.* 1998; Andreoli and Kayano 2006; Gonzalez *et al.* 2013). It is known that negative anomalies of SOI (El Niño) are associated with decreasing rainfall in northeastern Brazil. The study of SOI time series can provide guidance to understand the associatedin the atmospheric ity of theiated ocean/atmosphere coupling dynamics in the Pacific Ocean and its influence on the Walker and Hadley circulations. Changes in position and intensity of the Hadley and Walker cells are associated with changes in the large-scale atmospheric circulation patterns, which directly affects the meteorological conditions over the northeastern region of Brazil. This study aimed to interpret possible stationary signals, associated with a repetitive behavior in terms of long-range memory in SOI time series, as well as their influence on temporal variability of climate conditions in northeastern Brazil — Alcântara Launch Center (ALC).

## METHODOLOGY
### SOUTHERN OSCILLATION INDEX DATA

SOI data were obtained from the web page of the National Weather Service - Climate Prediction Center (www.cpc.ncep.noaa.gov/data/indices/). Monthly data normalized in the 1951–2015 period were used. The AutoSignal© software version 1.6 for Windows was employed to compute the autocorrelation function, which was calculated using the Fast Fourier Transform (FFT; Temperton 1985). The same software was used to estimate the Hurst Exponent ($H$) and generate graphs for analysis.

Figure 1 shows SOI time series analyzed to calculate $H$, using the rescaled range ($R/S$) analysis. $H$ is used to represent the behavior of time series, which presents persistence of features associated with a memory effect.

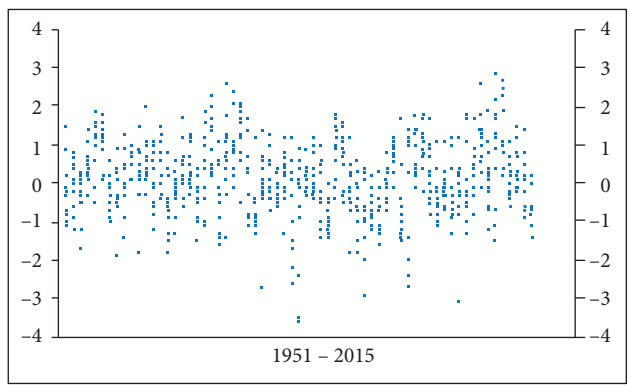

**Figure 1.** SOI from January 1951 to August 2015.

Table 1 presents statistical values of SOI time series. The raw data of this series were treated in Microsoft® Excel, in which the values of statistical information were generated. The coefficient of variations (CV), which is the standard deviation divided by the mean, is about 701.55%. The series features high dispersion relatively to the standard deviation. The first and second modes represent positions in SOI time series with greater frequency, characterizing the dynamic displacement between these 2 points within the time series. The statistical information in Table 1 indicates that SOI time series does not show a dynamic Gaussian Classic behavior.

**Table 1.** Statistical data of SOI series, with $n$ = 775 values.

| Statistic Parameters | Value |
|---|---|
| Mean | 0.135 |
| Median | 0.1 |
| First mode | 0.2 |
| Second mode | −0.1 |
| Standard deviation | 0.944 |
| Mean deviation | 0.738 |
| Variance | 0.892 |

# HURST EXPONENT — *R/S* METHOD

For a sequence $(X_t)^n_{t=1}$ of a time series, consider a partial sum $Y_k = \sum_{t=1}^{k} X_t$ — $1 \leq k \leq n$ and a sample variance $S_n^2 = (n-1)^{-1} \sum_{t=1}^{n} (X_t - X_n)^2$, where $X_n = n^{-1} \sum_{t=1}^{n} X_t$ is a sample mean. Rescaled adjusted range statistic *R/S* introduced by Hurst (1951) is defined as:

$$H = R/S = \frac{1}{S_n}\left\{\max_{1 \leq k \leq n}\left(Y_k - \frac{k}{n}Y_n\right) - \min_{1 \leq k \leq n}\left(Y_k - \frac{k}{n}Y_n\right)\right\} \quad (1)$$

Observe that the numerator Range (*R*) in Eq. 1 can be viewed as a range of partial sums of $X_t - X_n - t = 1, \ldots, n$ — or, equivalenttly, as the sum of the maximal and minimal distance of the partial sums $Y_k$ — $k = 1, \ldots, n$ — from a line passing through $Y_0 = 0$ and $Y_n$. The range (*R*) should be divided by the standard deviation (*S*) of the elements of time series to produce a standardized sequence *R/S* or resized value range. Thus, *R/S* is a measure of flutuations of the parcial sums of $(X_t)^n_{t=1}$ scaled by the standard deviation of observations.

The *H* estimate is a measure of long-term memory in time series. It relates to the autocorrelations of the time series and the rate at which the autocorrelation function decreases with the lag as the distance between pairs of values increases. An important aspect is that *H* serves as a measure of the fractal dimension of a data series. A white noise can be a discrete signal whose samples are considered as a sequence of random variables uncorrelated series with 0 mean and finite variance. Depending on the context, it may also require that these samples be independent and have the same probability distribution. In a particular situation, if each of the samples possess a normal distribution with a mean equal to 0, the signal is set to be Gaussian white noise.

A normally-distributed or Gaussian sequence may have a cumulative white noise and is known as regular Brownian motion or random walk (Kac 1947; van Horne and Parker 1967). The range or distance covering a variable in the normal Brownian motion will increase in proportion to the square root of time. To calculate the growth that may exist in a time series, a type of time-scaling ratio is used by the partitioning elements (number of observations) and generates an average of the other groups. The use of *R/S* analysis allows to obtain the statistic of fractal noise process and offers it as an alternative to the traditional Gaussian normal distribution.

The H variability of behavior can be assigned as the value calculated by the physical behavior of the analyzed time series; an *H* value equal to 0.5 would indicate no long-term memory. Higher *H* values indicate a growing presence of such an effect in the series. The duration of this memory effect is often visible as persistent and can be cyclic or not. An *H* value of 0.5, which features in its accumulated data series, is a random walk or pure Brownian motion. The dataset analyzed consists of true white noise in which each observation is completely independent of all previous observations, and the estimated autocorrelation series is essentially 0 everywhere, except at the 0 lag. Since *H* is less than 0.5, the temporal series have an anti-persistent behavior. Each value in the series tends to be more likely to have a negative correlation with the previous values.

These data series revert the signals more often than would be true for the white noise. Such systems are rare in geophysical time series. Much more common in nature are time series that present estimates of *H* values above 0.5. These characteristics in the behavior of the series, which are persistent, contain a memory effect. Therefore, each value of the series may be associated with a number of previous values of the same series. The modeling of autoregressive processes depends on exactly this purpose. For a persistent series, the series with autocorrelation tend to decrease their autocorrelation to 0. Both the analysis by estimating *H* and the correlation mapping show the memory effect on time series under review.

The hurstexp (x) function of the R package PRACMA version 3.2.2 (2015), from the R Foundation for Statistical Computing, calculates *H*. This relationship was derived from the MATLAB® code of Weron (2002), published in the MATLAB® central. This function returns a list of different definitions of *H* with different adjustments in their calculations, which are defined with the following components: $H_s$ — simplified approach $R/S$; $H_{rs}$ — simplified approach corrected $H$; $H_e$ — empirical $H$; $H_{al}$ — corrected empirical H; and $H_t$ —theoretical H. These different approaches are estimates of the H method, the corrected $R/S$ method and the corrected empirical method. The results are sometimes very different, depending on the series analysis in the study, which can be interpreted as estimates with highly-questionable values.

## FRACTAL DIMENSION

The random walks can be readily generalized to characterize fractal processes (Mandelbrot 1977, 1982) by introducing an additional parameter: $H$ ($0 < H < 1$). The variance is proportional to $\Delta t^{2H}$. In a fractal process, successive increments are correlated with coefficient of correlation ρ, independently of the time step $h$, where ρ is defined by the formula, using the moment technique of order 2 (Hastings and Sugihara 1993):

$$2^{2H} = 2 + 2\rho \quad (-0.5 < \rho < 1) \tag{2}$$

where: ρ = 0; $H$ = 0.5, which is a random process.

Mandelbrot (1983) and Goldberger (1996) created a definition of fractal dimension: it is an index that seeks to characterize patterns, whose order is fractal sets or the quantification of that nature in its complexity as a reason applied to change his own scale. The fractal dimension can be measured in 2 different ways; one of which is geometrically and the other is carried out by probability space. More useful to signal analysis is the definition of fractal dimension that uses probability space $1/H$. By this definition, a time series with a memory effect will have a fractal dimension between 1.0 and 2.0.

## LYAPUNOV EXPONENT

The R package TSERIESCHAOS version 3.2.2 (2015), from the R Foundation for Statistical Computing, is available on https://cran.r-project.org/web/views/TimeSeries.html. The 'tseriesChaos' algorithms generate analysis for non-linear time series. These algorithms were developed by Di Narzo (2013) and are available on https://cran.r-project.org/web/packages/tseriesChaos/index.html (Hegger et al. 1999; Rosenstein et al. 1993). Two functions are used: function lyap_k ($\lambda_1$) estimates the largest Lyapunov exponent of a given scalar time series using the algorithm of Kantz (Hegger et al. 1999) and function lyap computes the Lyapunov regression coefficients for the time series segment given as input, in this case, SOI data. Tools to evaluate the maximal Lyapunov exponent of a dynamic system from a univariate time series provide a parameter that characterizes the dynamics of an attractor. It measures the rate of divergence of neighboring orbits within the attractor and thus quantifies the dependence or system sensitivity to initial conditions. The existence of at least 1 positive Lyapunov exponent is a strong indication of the presence of chaos in the system.

## PERMUTATION TEST

The permutation test is used to analyze the relationship between the signals in the temporal series of SOI and the monthly means of surface wind and monthly means of maximum surface wind, from January 1951 to December 1999 (Good 2005; Collingridge 2013). The wind data used were from the São Luís Airport near the ALC. By definition, the vector **P** is the wind monthly mean and **J** ($N \times 1$) is the monthly value of SOI. The test seeks to make random permutations of **J**, keeping **P** fixed. For each permutation, the correlation between vectors **P** and **J** was calculated, resampling the series in the order of 10,000 times, thus building the distribution of correlations (*r*). From these distributions, the value that represents the confidence interval at the 5% level of the correlations in the upper or lower tail of distributions (*r* critical) can be obtained. The critical value of the correlation at 10,000 times is permuted, ordered from the smallest to the largest value; the calculated value at the 9,500º position is the critical one at the 5% level in the reconstructed distribution (in the upper tail) or in the lower tail; the critical value is that in the 500º position. The time series contains 587 monthly values of wind average speed and maximum wind average values. The permutation test method is statistically more robust than the classical one to test the correlation signal between different series that have low correlation values.

## RESULTS

Figure 2 shows the behavior of the autocorrelation function, in which the time series decays to 0 for a 13-month period (lag 13). The autocorrelation function exhibits positive correlation of about 0.14 for the return times (persistence characteristics) between 25 and 40 lag (2 –3 years) and between 52 and 64 lag (5 – 6 years), with higher significance for the latter. In the analysis of autocorrelation function in a time series, with a magnitude of the order of 0.14, this value should not be interpreted as a simple randomness in the time series but as a value of relative significance. It may represent possible temporal scale influences with periods of 2 – 3 and 5 – 6 years, respectively. The results were similar to those obtained by An and Wang (2000) using another method: wavelet analysis. They showed that the period of oscillation has increased from 2 – 4 years (high frequency), in 1962 – 1975, to 4 – 6 years (low frequency), in 1980 – 1993.

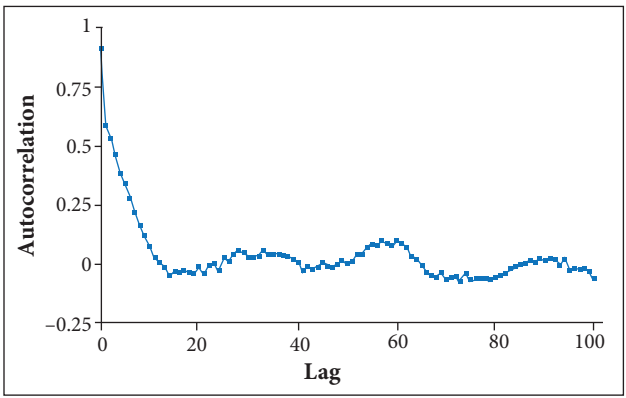

**Figure 2.** Series of autocorrelation of SOI values between 1951 and 2015.

Jin et al. (1996), analyzing ENSO and the existence of annual and sub-harmonic cycles of frequency block and aperiodicity, observed the transition and the occurrence of chaotic regimes close to 4 years and with a quasi-biennial peak, being produced by a dynamic of non-linear interactions.

In the study of Chang et al. (1995), on the interactions between the seasonal cycle and ENSO in an intermediate coupled Ocean-Atmosphere model, the ENSO cycle falls within a sandwiched regime between 3- and 2-year frequency-locking regimes. When the existence of a strange attractor for SOI with a fractal dimension of 2.5 to 6 (with a value close to 5.2) had been estimated, the analysis was performed by a simulation of 1,000 years using the values of monthly SST time series. The observations in their study suggest a change of frequency, and this was accompanied by a significant change in the structure of coupled ENSO mode. This result shows the trend of persistency in decadal time series.

Figure 3 shows the statistic range and suggests a positive long-range autocorrelation. The distribution shows persistent or highly-periodic behavior. The results for $H$ calculation were: $H = 0.561$; $S = 0.013$; fractal dimension $1/H = 1.78$ and coefficient of determination $R^2 = 1 - (SSE/SSM) = 0.911$, where SSE is the sum of squared errors (residuals) and SSM is the sum of squares around the mean; weighted $H = 0.704$; weighted $S = 0.007$; weighted fractal dimension $1/H = 1.420$ and $R^2$, weighted by $S$, was: $R^2 = 1 - (SSE/SSM) = 0.998$. The coefficient of correlation $\rho = 0.088$, in the time series, has a fractal correlation value of about 10%, and the coefficient of correlation weighted by $S$ is 0.326.

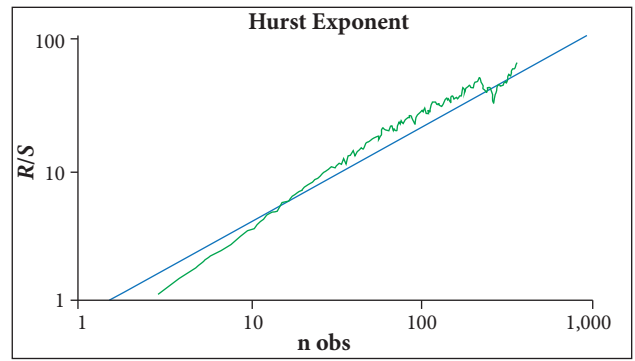

**Figure 3.** $H$ of SOI time series calculation between 1951 and 2015, normalized by $S$.

SOI time series with the $H$ calculation presents a long-range memory with persistence, with values of $H = 0.561$ and $1/H$ of about 1.78 with fractal dimension. These results can be interpreted as the relative tendency of a time series to strongly regress to its mean or to be grouped in 1 direction (Kleinow 2002). The results in Table 2 show that the $H$ estimates using the function hurstexp (x) in R produce values above 0.5, characterizing a time series with long-range memory.

**Table 2.** Results of the function hurstexp (x) from R package.

| Hurst exponent | Value |
|---|---|
| Simple $R/S$ Hurst estimation | 0.694 |
| Corrected $R$ over $S$ Hurst exponent | 0.775 |
| Empirical $H$ | 0.762 |
| Corrected empirical $H$ | 0.739 |
| Theoretical $H$ | 0.533 |

Figure 4 shows the result of the function lyap_k with positive values in different time steps of SOI time series. The function lyap estimated with Lyapunov regression coefficient for SOI time series is $Ƙ_1 = 0.40$. These results show that SOI time series presents a chaotic behavior, which is consistent with Chang et al.

(1995), who propose that the ENSO irregularity can be viewed as a chaotic low-order process driven by the seasonal cycle. The main characteristic of the behavior of chaotic systems is a high sensitivity in the initial conditions, which implies that the evolution of the system can be modified by small disturbances in the system.

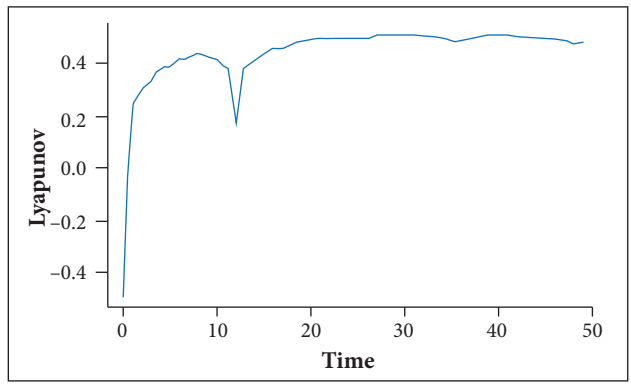

**Figure 4.** The largest Lyapunov exponent for SOI time series.

Figure 5 shows the result of the permutation test between SOI and the resulting monthly mean of zonal and meridional wind components from São Luís Airport surface station in the Maranhão State. The calculated result shows statistical evidence that the temporal series has a significant correlation with the 5% level of confidence interval. The correlation calculated (−0.230) is more displaced in the lower position of the tail of the distribution than the critical $r$ value (−0.068) in the lower tail. The monthly mean wind can be associated with SOI variation signal, as the correlation was negative, showing that, in the months when SOI is negative (situation with El Niño), wind average would be higher.

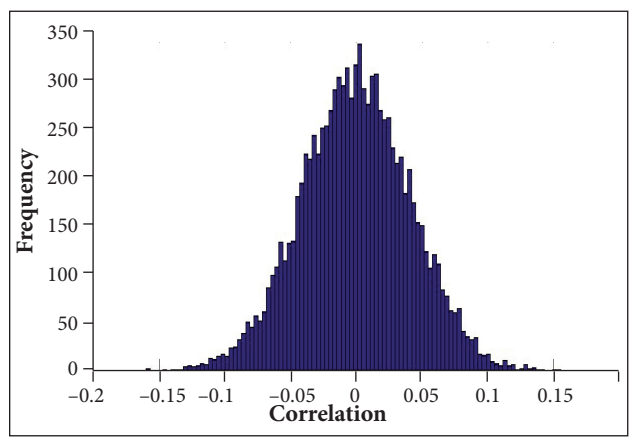

**Figure 5.** Distribution of correlations between the monthly average of the wind and SOI.

In the years of El Niño occurrence (negative SOI), the characteristics of dynamic systems in the equatorial region (the Hadley and Walker circulations) presented significant alterations, and their effects influence atmospheric behavior in the Brazilian northeast. Operationally, this would influence directly the ALC, with greater intensity to the wind profile near the surface.

Marques and Oyama (2015) conducted a study on the interannual variability of precipitation in ENSO-neutral years for the ALC. Using various gridded datasets for the 1951 – 2010 period, the authors observed that, below average precipitation, it was related to strong east-norteasterly low-level winds (925 hPa) and northward of the interhemispheric gradient of the sea surface temperature anomalies over Atlantic (GRAD), as well as that, for El Niño conditions, northward GRAD would intensify the negative precipitation anomalies in the northern-northeastern Brazil.

Figure 6 presents, in a similar manner, the test shown in Fig. 4, but the results show the correlation between SOI series and the monthly average of the wind maximums. The results statistically showed a correlation between the signals of the time series with a significant confidence interval at the 5% level. The averages of the monthly maximum winds would be higher in months with negative SOI (El Niño). The calculated correlation (−0.191) is displaced in the lower position of the tail of the distribution than the critical $r$ value (−0.069).

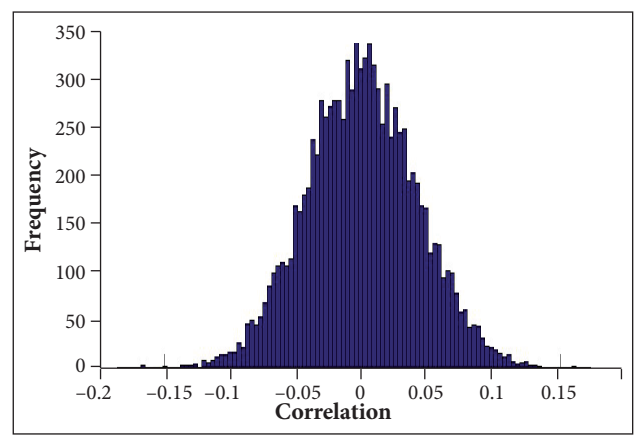

**Figure 6.** Distribution of correlations between the monthly average maximum wind and SOI.

## DISCUSSION

The application of the $H$ methodology can capture the trends of SOI time series with significant results. This indicates that the *H* method can be used as an alternative technique, allowing to show that SOI time series presents a long-range memory which is persistent throughout the lag range series. Using FFT, a significant signal was obtained, suggesting a persistence in the time series. This methodology has shown strong evidence that the *H* estimative is above 0.5, as the *H* value can represent a measure of the fractal dimension. A low fractal dimension presents greater coherence in time series and is more predictable; on the other hand, with high fractal dimension, it will have a behavior with less predictability.

The *H* calculation allows in its methodology to estimate a fractal dimension of SOI time series of about 1.78. This result can indicate chaotic behavior in SOI time series, thus quantifying the dependency or the sensitivity of the system to the initial ENSO conditions. Also, the ENSO irregularity can be viewed as a chaotic low-order process driven by the seasonal cycle.

## CONCLUNDING REMARKS

The results show the importance of Ocean-atmosphere interaction, where there are non-linear interactions. Associated teleconnections effects create complex dynamics, influencing and modulating the climatic indexes such as SOI.

In El Niño occurrence (negative SOI), there is a statistical correlation with the monthly mean wind and its maximum monthly mean, as well as a tendency to influence the average wind profile behavior near the surface, which operationally affects the ALC. El Niño characterizes a situation with stronger winds and predominant direction (east-northeasterly). Northward GRAD would intensify the negative precipitation anomalies in the northern-northeast Brazil.

The series features long-range memory behavior; therefore, this preliminary analysis indicates that autoregressive models can be used in the seasonal forecast for the Brazilian northeast and for trend studies with interest in seasonal variability range for the ALC. The knowledge of the SOI behavior and its association with wind variability allows a better prediction of wind intensity and, consequently, an improvement in the safety of ALC launch activities.

## ACKNOWLEDGEMENTS

The authors thank the support of the Instituto de Aeronáutica e Espaço (IAE).

## AUTHOR'S CONTRIBUTION

Conceptualization and Methodology, Corrêa CS. and Schuch DA; Writing – Review & Editing, Corrêa CS, Queiroz AP, Fisch G, Corrêa FN, and Coutinho MM.